\documentclass[pra,twocolumn,superscriptaddress]{revtex4} 

\usepackage{graphicx}
\usepackage{bm}
\usepackage{amsmath}
\usepackage{hyperref}
\usepackage{amssymb}
\usepackage{amsfonts}
\usepackage{fancyhdr}
\usepackage{slashed}
\usepackage{latexsym,epsfig,bbm}
\usepackage{mathrsfs}
\usepackage{setspace}

\usepackage{comment}

\usepackage{braket}

\usepackage{color}
\definecolor{blue}{rgb}{0,0.2,1}

\definecolor{red}{rgb}{0.9,0,0}

\begin{document}

\title{Quantum Hopfield neural network}

\author{Patrick Rebentrost}
\email{pr@patrickre.com}
\affiliation{Xanadu, 372 Richmond Street W, Toronto, Ontario M5V 1X6, Canada}

\author{Thomas R. Bromley}
\email{tom@xanadu.ai}
\affiliation{Xanadu, 372 Richmond Street W, Toronto, Ontario M5V 1X6, Canada}

\author{Christian Weedbrook}
\affiliation{Xanadu, 372 Richmond Street W, Toronto, Ontario M5V 1X6, Canada}

\author{Seth Lloyd}
\affiliation{Massachusetts Institute of Technology, Department of Mechanical Engineering,  77 Massachusetts Avenue, Cambridge, Massachusetts 02139, USA}

\date{\today}

\begin{abstract}
Quantum computing allows for the potential of significant advancements in both the speed and the capacity of widely-used machine learning techniques. Here we employ quantum algorithms for the Hopfield network, which can be used for pattern recognition, reconstruction, and optimization as a realization of a content addressable memory system. We show that an exponentially large network can be stored in a polynomial number of quantum bits by encoding the network into the amplitudes of quantum states. By introducing a new classical technique for operating the Hopfield network, we can leverage quantum algorithms to obtain a quantum computational complexity that is logarithmic in the dimension of the data. 
We also present an application of our method as a genetic sequence recognizer.
\end{abstract}

\maketitle 

\section{INTRODUCTION}

Machine learning is an interdisciplinary approach that brings together the fields of computer science, mathematics, statistics, and neuroscience with the objective of giving computers the ability to make predictions and generalizations from data~\cite{bishop2006pattern}. A typical machine learning problem falls into three main categories: supervised learning, where the computer learns from a set of training data; unsupervised learning, with the objective of identifying underlying patterns in data; and reinforcement learning, where the computer evolves its approach based on real-time feedback. Machine learning is changing how we interact with technology in areas such as autonomous vehicles, the internet of things, and e-commerce.

Quantum information science has developed from the idea that quantum mechanics can provide improvements in information processing and communication~\cite{nielsen2002quantum}. The promises of quantum information are manifold, ranging from exponentially fast quantum computers
, information theoretic secure quantum communication networks, to high precision measurements useful in science and technology
. Over the past few decades, quantum information science has transitioned from scientific theory to a viable form of technology. 

Given the encouraging technological implications of both machine learning and quantum information science, it was inevitable that their paths would crossover to form quantum machine learning~\cite{schuld2015introduction,biamonte2016quantum,ciliberto2017quantum,dunjko2017machine}
. Quantum-enhanced machine learning approaches use a toolbox of quantum subroutines to achieve computational speed-ups for established machine learning algorithms. This toolbox includes fundamentals like quantum basic linear algebra subroutines (qBLAS), including eigenvalue finding~\cite{
nielsen2002quantum}, matrix multiplication~\cite{wiebe2012quantum} and matrix inversion~\cite{harrow2009quantum}
. 
One can also build on quantum techniques, such as amplitude amplification~\cite{brassard2002quantum,grover1996fast}
and quantum annealing~\cite{kadowaki1998quantum,finnila1994quantum,boixo2013quantum}.
These elements have been put together in recent works on quantum machine learning~\cite{wiebe2014quantum,dunjko2016quantum,benedetti2016estimation,romero2016quantum,schuld2014quantum,zhao2015quantum,wossnig2017quantum}
, including nearest-neighbor clustering~\cite{wiebe2015quantum}, the quantum support vector machine~\cite{rebentrost2014quantum}, and quantum principal component analysis~\cite{lloyd2013quantum,kimmel2017hamiltonian}. 

Artificial neural networks 
are highly successful in machine learning and are hence of special interest for quantum adaptation~\cite{schuld2014quest,wiebe2014quantum,amin2016quantum,benedetti2017quantum,romero2016quantum}. A collection of binary or continuous-valued neurons are connected and evolve in such a way that each neuron decides its state based upon a weighted function of the neurons connecting to it. The neurons can be organized into layers and may be configured to allow for backflow of information (known as a recurrent network, often constructed from building blocks of long short-term memory~\cite{hochreiter1997long}). We focus on the Hopfield network, which is a single layer, recurrent and fully connected neural network with undirected connections between neurons. Such networks can be trained using the Hebbian learning rule~\cite{hebb1949organization}, based on the notion that the connection weights are stronger when they are regularly fired together from training data. The Hopfield network can act as a non-sequential associative memory, with technological application in image processing and optimization~\cite{
cheng1996application} 
 and wider interest in neuroscience and medicine.

State of the art  neural networks are based on deep learning methods with many hidden layers and using learning rules such as stochastic gradient descent~\cite{Hinton06,Bengio09}. While the Hopfield network is not competitive with these modern neural networks, it is interesting to investigate the \emph{quantum} context for several reasons. The fully visible structure allows a simple encoding of the information into the amplitudes of a quantum state. With such an encoding, techniques such as quantum phase estimation and matrix inversion can be applied which have exponentially fast run times in certain cases.
Learning rules such as Hebbian learning find a relatively straightforward representation in the quantum domain. Finally, Hebbian learning and the Hopfield network were one of the early neural networks methods and fast quantum algorithms are interesting as building blocks for more advanced quantum networks.

We present in this article a method to construct a quantum version of the Hopfield network (qHop), resulting from a new adaptation of the classical Hopfield network when specialized to the situation of information erasure. 
The network state is embedded into the amplitudes of a quantum system composed of a register of quantum bits (qubits). 
Our approach differs from previous generalizations of the Hopfield network; Refs.~\cite{akazawa2000quantum,behrman2006microtubules} focussed on the condensed matter/biology setting, Ref.~\cite{rotondo2017open} encoded neurons directly into qubits, Ref.~\cite{
ventura1998quantum} used a quantum search, while Ref.~\cite{seddiqi2014adiabatic} 
 harnessed quantum annealing. 
 The training of qHop is here addressed by introducing quantum Hebbian learning, whereby the symmetric graph weighting matrix can be associated to a density matrix stored in a qubit register. We show how this density matrix can be used operationally to imprint relevant training information onto the system. The next step is to operate qHop efficiently. To this end, we propose a new approach to optimizing the classical Hopfield network using matrix inversion. Matrix inversion can under certain conditions be performed efficiently using quantum algorithms with a run time $\mathcal{O}({\rm poly}\left(\log d\right))$ in the size of the matrix
$d$~\cite{harrow2009quantum}
. By combining these algorithms with the quantum Hebbian learning subroutine and sparse Hamiltonian simulation~\cite{berry2015hamiltonian}
, we formalize our algorithm qHop. Using qHop can therefore provide speedups in the application of the Hopfield network as a content addressable memory system. As an example application, we consider the problem of RNA sequence pattern recognition of the influenza A virus in genetics. We use this scenario to compare the recovery performances of both approaches to operating the Hopfield network.


\section{Neural networks}

Let us first outline some basic features of neural networks. Consider a collection of $d$ artificial binary-valued neurons $x_{i} \in \{1,-1\}$ with $i \in \{1, 2, \ldots , d\}$~\cite{mcculloch1943logical}, that are together described by the activation pattern vector $\bm{x} = \{x_{1}, x_{2}, \ldots ,x_{d}\}^{\intercal}$, with $\bm{x}^{\intercal}$ denoting the transpose of $\bm{x}$. The neurons are formed into a (potentially multilayer) network by wiring them to create a connected graph, which can be specified by a real and square $(d \times d)$-dimensional weighting matrix $W$. Its elements $w_{ij}$ specify the neuronal connection strength between neurons $i$ and $j$~\cite{hopfield1982neural}. We note that each neuron is not typically self-connected, so that $w_{ii} = 0$. Furthermore, for an undirected network, $W$ is symmetric. In addition, we may also use continuously activated neurons in both classical and quantum settings, but focus in this work on the binary case for the input and test patterns.

Setting the weight matrix $W$ is achieved by teaching the network a set of training data. This training data can consist of known activation patterns for the visible neurons, i.e. the input and output neurons, with the learning achieved using tools such as backpropagation, gradient descent and Hebbian learning
. A network can be fully visible, so that every neuron acts as both an input and an output.

The Hopfield network is a single layered, fully visible, and undirected neural network. Here, one can teach the network using the Hebbian learning rule~\cite{hebb1949organization}. This rule sets the weighting matrix elements $w_{ij}$ according to the number of occasions in the training set that the neurons $i$ and $j$ fire together
. Consider a training set of $M$ activation patterns $\bm{x}^{(m)}$, with $m \in \{1,2,\ldots,M\}$. The (normalized) weighting matrix is given by
\begin{equation}\label{Eq:WeightingMatrix}
W = \frac{1}{M d}  \left[ \sum_{m=1}^{M}\bm{x}^{(m)}\left(\bm{x}^{(m)}\right)^{\intercal}\right] - \frac{\mathbb{I}_{d}}{d}  ,
\end{equation}
with $\mathbb{I}_{d}$ the $d$-dimensional identity matrix.

\section
{Quantum neural networks}

Now we consider the task of using multi-qubit quantum systems to construct quantum neural networks. One established method is to have a direct association between neurons and qubits~\cite{schuld2014quest}, unlocking access to quantum properties of entanglement and coherence. We instead encode the neural network into the amplitudes of a quantum state. This is achieved by introducing an association rule between activation patterns of the neural network and pure states of a quantum system. Consider any $d$-dimensional vector $\bm{x} := \{x_{1},x_{2},\ldots,x_{d}\}^{\intercal}$. We associate it to the pure state $\ket{x}$ of a $d$-level quantum system according to $\bm{x} \rightarrow \left \vert \bm{x}\right \vert_2 \ket{x}$, with $\left|\bm{x}\right|_2 = \sqrt{\sum_{i=1}^{d}x_{i}^{2}}$ the $l_{2}$-norm of $x$ and
$\ket{x} := \frac{1}{\left|\bm{x}\right|_2} \sum_{i=1}^{d} x_{i} \ket{i}$
written with respect to the standard basis such that $\braket{x|x} = 1$. Note that for activation pattern vectors with $x_i=\pm 1$, the normalization is $|\bm x|^2_2 = d$. The $d$-level quantum system can be implemented by a register of $N = \lceil \log_{2} d \rceil$ qubits, so that the qubit overhead of representing such a network scales logarithmically with the number of neurons. We discuss in the following section how the weighting matrix $W$ can be understood in the quantum setting by using quantum Hebbian learning.

Crucial for quantum adaptations of neural networks is the classical-to-quantum read-in of activation patterns. In our setting, reading in an activation pattern $\bm{x}$ amounts to preparing the quantum state $\ket{x}$. This could in principle be achieved using the developing techniques of quantum random access memory (qRAM)~\cite{giovannetti2008quantum} or efficient quantum state preparation, for which restricted, oracle based, results exist~\cite{soklakov2006efficient}
. In both cases, the computational overhead can be logarithmic in terms of $d$. 
State preparation routines can potentially be made more robust by the insight that certain errors can be tolerated in the machine learning setting \cite{Zhao2018}.
One can alternatively adapt a fully quantum perspective and take the activation patterns $\ket x$ directly from a quantum device or as the output of a quantum channel. For the former, our preparation run time is efficient whenever the quantum device is composed of a number of gates scaling at most polynomially with the number of qubits. Instead, for the latter, we typically view the channel as some form of fixed system-environment interaction that does not require a computational overhead to implement.

\section{Quantum Hebbian learning}
\label{secQHeb}

Using our association rule, the training set of activation patterns $\bm{x}^{(m)}$ can be associated with an ensemble of pure quantum states $\ket{x^{(m)}}$. Let us now focus on the Hopfield network, with a weighting matrix $W$. We first introduce the \emph{quantum} Hebbian learning algorithm (qHeb), which relies on two important insights: (i) that one can associate the weighting matrix $W$ directly to a mixed state $\rho$ of a memory register of $N$ qubits according to
\begin{equation}
\rho := W +\frac{\mathbb{I}_{d}}{d}  
=\frac{1}{M}\sum_{m=1}^{M} \ket{x^{(m)}}\bra{x^{(m)}},
\end{equation}
and (ii), one can efficiently perform quantum algorithms that harness the information contained in $W$.

To comment on (i), the problem of efficient preparation of $\ket{x^{(m)}}$ can be addressed using any of the techniques discussed in the previous section. We denote by $T_{\rm in}$ the required run time to prepare each $\ket{x^{(m)}}$. In the situations discussed above $T_{\rm in}\in \mathcal{O}\left({\rm poly}\left( \log d\right)\right)$.

Regarding (ii), now suppose that we have prepared $\rho$ in the laboratory and want to harness the training information contained within. If $\rho$ is the direct output of an unknown quantum device, then we cannot recover the training states $\ket{x^{(m)}}$, since the decomposition of $\rho$ into pure states is not unique. On the other hand, we can still obtain useful information about $\rho$, such as its eigenvalues and eigenstates. One approach to do this could be to perform a full quantum state tomography of $\rho$. For states with low rank $r$, there exists tomographical techniques with a run time $\mathcal{O}\left({\rm poly}\left(d \log d, r \right)\right)$~\cite{gross2010quantum}, although for some cases the required run time for full state tomography can grow polynomially with the number of qubits~\cite{cramer2011efficient}. 

We show that one can use $\rho$ as a ``quantum software state''~\cite{kimmel2017hamiltonian}. That is, it is possible to efficiently simulate $e^{i \rho t}$ for time $t$ to precision $\epsilon$ with a required run time approximately $T_{\rm qHeb} \in \mathcal{O}\left({\rm poly}\left(\log d ,t , M,\frac{1}{\epsilon} \right)\right)$. One can then use this ability to estimate the eigenvalues  and eigenstates of $\rho$ to precision $\epsilon$ through the quantum phase estimation algorithm~\cite{
nielsen2002quantum}, requiring an overall run time $T_{\rm eigenvalues} \in \mathcal{O}\left({\rm poly}\left(\log d , \frac{1}{\epsilon},M\right)\right)$.

Let us define the set of $M$ unitary operators $\{\mathcal{U}_{k}\}_{k=1}^{M}$ acting on an $N+1$ register of qubits according to
\begin{equation}\label{Eq:ConditionalSimulation}
\mathcal{U}_k := \ket 0 \bra 0 \otimes \mathbbm I + \ket 1 \bra 1 \otimes e^{-i  \ket{ x^{(k)}} \bra {x^{(k)}}  \Delta t}.
\end{equation}
The unitaries apply the different memory pattern projectors  $\ket{ x^{(k)}} \bra {x^{(k)}} $ conditionally and for a small time $\Delta t$.
We now show how to simulate these unitaries and that one can simulate a conditional $e^{-i \rho t}$ by applying them for a suitably large number of times. 
Let $S$ be the swap matrix between the subsystems for $\sigma$ and $\ket{ x^{(k)}}$. 
Note that 
\begin{eqnarray}
\mathcal{U}_S &:=& e^{-i \ket 1 \bra 1 \otimes S \Delta t} \nonumber \\ &=& \ket 0 \bra 0 \otimes \mathbbm{I}+ \ket 1 \bra 1 \otimes e^{-i S \Delta t},
\end{eqnarray}
where $\ket 1 \bra 1 \otimes S $ is $1$-sparse and efficiently simulatable.
For sparse Hamiltonian simulation, the methods in Ref.~\cite{
berry2007efficient,berry2015hamiltonian}
can be used
with a constant number of oracle calls and run time $ \tilde{\mathcal{O}}(\log d)$, where we omit polylogarithmic factors in $\mathcal{O}$
by use of the symbol $\tilde{\mathcal{O}}$. Note that 
\begin{eqnarray}
{\rm tr}_2 \left \{\mathcal{U}_S \left( \ \ket q \bra q \otimes \ket{ x^{(k)}} \bra{x^{(k)}} \otimes \sigma\right) \ \mathcal{U}_S^\dagger \right \} \nonumber \\ = \mathcal{U}_k \,  \left( \ket q \bra q \otimes \sigma\right) \, \mathcal{U}_k^\dagger + \mathcal{O}(\Delta t^2).
\end{eqnarray}
The trace is over the second subsystem containing the state $\ket{ x^{(k)}}$. 
Thus the subsystem of ancilla qubit and $\sigma$ effectively undergoes time evolution with  $\mathcal{U}_k$. 

We now apply the $M$ unitaries $\mathcal{U}_{k}$ sequentially for $n$ repetitions. i.e.~we perform
\begin{equation}
U_{t}:=\left(\prod_{k=1}^{M} \mathcal{U}_{k}\right)^{n} 
\end{equation}
with $\Delta t =  t/nM$. Consider for the sake of simplicity the unconditioned evolution. Using the standard Suzuki-Trotter method~\cite{Childs2017}, it follows that
\begin{eqnarray}\label{eqHebbianLieProduct}
\epsilon &:=& \left \Vert \left(e^{-i  \ket{ x^{(1)}} \bra {x^{(1)}}  t/(nM)}\ldots e^{-i  \ket{ x^{(M)}} \bra {x^{(M)}}  t/(nM)}\right)^{n} \right. \nonumber \\ & & \qquad \qquad \left.  - e^{-i  \rho t}\right \Vert \in \mathcal{O}\left( \frac{ t^{2}}{n} \right).
\end{eqnarray}
Hence, we require $n \in \mathcal{O}\left(\frac{ t^{2}}{\epsilon}\right)$ repetitions, with each repetition requiring $M$ sparse Hamiltonian simulations. This results in a run time $\mathcal{O}\left(\frac{M t^{2}}{\epsilon} \right)$. The advantages of this approach is that we can use copies of the training states $\ket{x^{(m)}}$ as ``quantum software states" \cite{kimmel2017hamiltonian} and, in addition, we do not require superpositions of the training states.
In summary, 
we can simulate $\rho$ conditionally to a precision $\epsilon$ with a number of applications of $\mathcal U_k$ of order $\mathcal{O}\left(M t^2/ \epsilon\right )$. Each $\mathcal{U}_k$ can be realized with logarithmic run time using sparse Hamiltonian simulation~\cite{
berry2015hamiltonian}, resulting in the overall run time of $T_{\rm qHeb} \in \mathcal{O}\left({\rm poly}\left(\log d ,t , M,\frac{1}{\epsilon} \right)\right)$.
 
The quantum phase estimation algorithm~\cite{
nielsen2002quantum,harrow2009quantum} can then be implemented to find the eigenvalues $\mu_{j}(\rho)$ and corresponding eigenstates $\ket{v_{j}(\rho)}$ of $\rho$. Here we prepare a register of $T$ qubits additional to our register of $N$ qubits in the composite state $\sum_{t=1}^{2^{T}} \ket t \otimes \ket \psi$ for some arbitrary $\ket{\psi}$. The size of $T$ is set by the precision with which we wish to estimate the eigenvalues. Applying the controlled unitaries 
$U_t$ results in the state $\sum_j \beta_j \ket{\tilde \mu_{j}(\rho)} \otimes \ket {v_{j}(\rho)}$. Each $\ket{\tilde \mu_{j}(\rho)}$ contains an approximation of the eigenvalues $\mu_{j}(\rho)$~\cite{nielsen2002quantum}, and $\beta_j := \bra{v_{j}(\rho)} \psi \rangle$. If we take $2^T \in\mathcal{O}\left(1/\epsilon\right)$, we can estimate the eigenvalues of $\rho$ to precision $\epsilon$ with a number of copies of the memory states $\ket{x^{(m)}}$ of the order
$\mathcal{O}\left(M/\epsilon^{3}\right)$. This results in an overall run time $T_{\rm eigenvalues} \in \mathcal{O}\left({\rm poly}\left(\log d , \frac{1}{\epsilon}, M  \right)\right)$. Our quantum Hebbian learning method thus shows how to prepare the weight matrix from the training data as a mixed quantum state and then specifies how that density matrix can be used in a quantum algorithm for higher-level machine-cognitive function, specifically to learn eigenvalues and eigenvectors. 

\begin{figure*}
  \begin{center}
  \includegraphics[width=\textwidth]{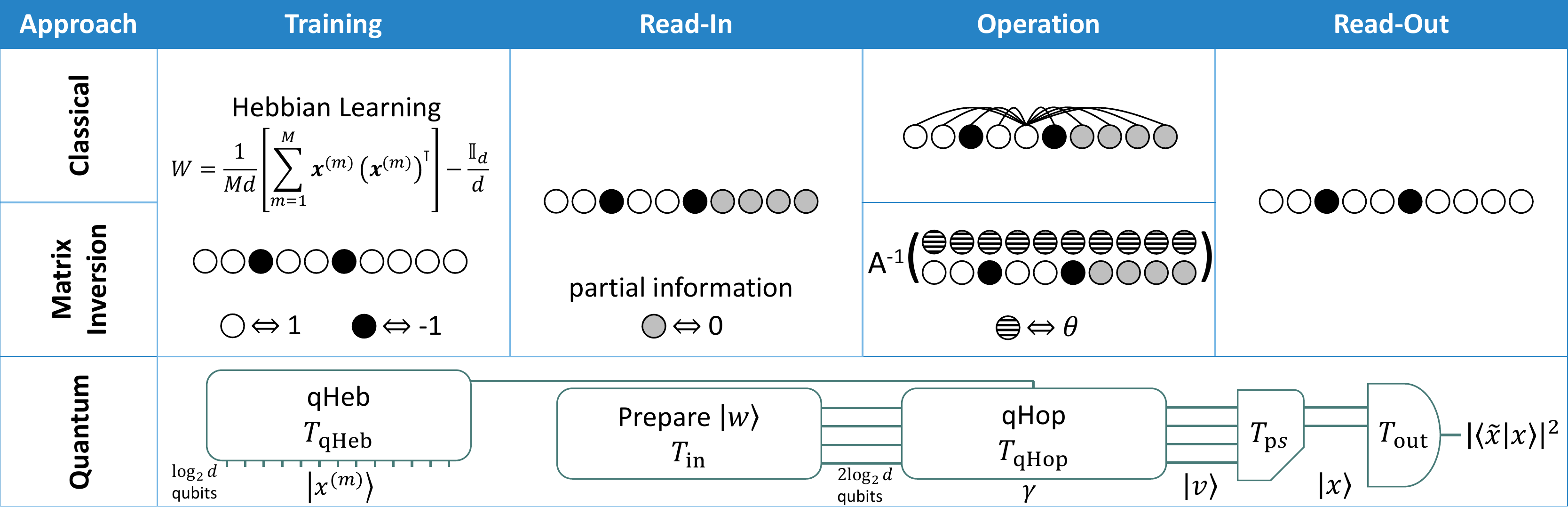}
\end{center}
    \vspace{-1em}
\caption{
The Classical and Quantum Hopfield networks. We discuss three approaches to operating the network. The standard classical approach is to iteratively update the neurons based on the connections to neighboring neurons. Our newly-developed classical approach solves a relaxation of the problem posed by a linear equation system and solvable through matrix inversion. Hebbian learning is employed to set the weighting matrix $W$ from $d$-length training data $\{\mathbf{x}^{(m)}\}_{m=1}^{M}$. The third approach uses qHop, encoding data in order $\log_{2}d$ qubits. Here, the pure state $\ket{w}$ is first prepared which contains user-defined neuron thresholds and a partial memory pattern. Our qHop algorithm proceeds to calculate $\ket{v} = A^{-1}\ket{w}$, with the matrix $A$ containing information on the training data and regularization $\gamma$. To achieve this, we introduce the quantum Hebbian learning algorithm qHeb for density matrix exponentiation of the mixture $\rho$ detailing training data $\ket{x^{(m)}}$. The output pure state $\ket{v}$ contains information on the reconstructed state $\ket{x}$ and Lagrange multipliers, which are post-selected out. The result $\ket{x}$ can be accessed through global properties such as the swap test, which uses multiple copies of $\ket{x}$ to measure the fidelity $\left|\braket{\tilde{x}|x}\right|^{2}$ with another state $\ket{\tilde{x}}$. The required run time for each step is given by the subscripted $T$.
}
  \label{Fig:Hopfield}
\end{figure*}

\section{The Hopfield network}

We return to the classical Hopfield network and discuss its operation, having already shown the Hebbian learning rule to store $M$ activation patterns in the weighting matrix $W$, see also Fig.~\ref{Fig:Hopfield} for a diagram.
Suppose that we are supplied with a new activation pattern, $\bm{x}^{({\rm new})}$, in the form of a noise-degraded  version of one from the training set or alternatively a similar pattern that is to be compared to the training set. In the following, we show the standard way of operating the network and then develop a new method based on matrix inversion.

The standard method of operating the Hopfield network proceeds by initializing it in the activation $\bm{x}^{({\rm new})}$ and then running an iterative process whereby neuron $i$ is selected at random and updated according to the rule
\begin{equation}\label{Eq:ActivationRule}
x_{i} \rightarrow \left\lbrace \begin{array}{ll}
+1 \qquad & {\rm if} \,\,\,\, \sum_{j=1}^{d} w_{ij}x_{j} \geq \theta_{i} \\
-1 & {\rm otherwise,}
\end{array}\right.
\end{equation}
with $\bm{\theta} := \{\theta_{i}\}_{i=1}^{d}\in \mathbb{R}^{d}$ a user-specified neuronal threshold vector that determines the switching threshold for each neuron. Each element $\theta_{i}$ should be set so that its magnitude is of order at most $1$. The result of every update is a non-increase of the network energy
\begin{equation}\label{Eq:HopfieldEnergy}
E = - \frac{1}{2}\bm{x}^{\intercal}W \bm{x} + \bm{\theta}^{\intercal}\bm{x},
\end{equation}
with the network eventually converging to a local minimum of $E$ after a large number of iterations.

Since $W$ has been fixed due to the Hebbian learning rule so that each $\bm{x}^{(m)}$ is a local minimum of the energy, the output of the Hopfield network is ideally one of the trained activation patterns. The utility of such a memory system is clear and the Hopfield network has been directly employed, for example, in imaging~\cite{
cheng1996application}.

We now introduce another approach to operating the classical Hopfield network, see Fig.~\ref{Fig:Hopfield}. Suppose that we are supplied with incomplete data on a neuronal activation pattern such that we only know the values of $l < d$ neurons with labels $\mathcal{L} \subset \{1,2,\ldots,d\}$. This setting corresponds to noise-free information erasure. We can initialize our activation pattern to be $\bm{x}^{\rm (inc)}:=\{x_{1}^{{\rm (inc)}},x_{2}^{{\rm (inc)}},\ldots,x_{d}^{{\rm (inc)}}\}^{\intercal}$ with $x_{i}^{{\rm (inc)}} = x_{i}^{{\rm (new)}}$ if $i \in \mathcal{L}$ and $x_{i}^{{\rm (inc)}} = 0$ otherwise. Our objective is to use the trained Hopfield network to recover the original activation pattern $\bm{x}^{\rm (new)}$. An alternative use of the Hopfield network when supplied with a noisy new pattern is shown in Appendix \ref{appendixNoisy}.

Let us first define the projector $P$ onto the subspace of known neurons, such that $P$ is diagonal with respect to the standard basis. We proceed by minimizing the energy $E$ in Eq.~(\ref{Eq:HopfieldEnergy}) subject to the constraint that $P \bm{x} = \bm{x}^{({\rm inc})}$. The Lagrangian for this optimization is
\begin{equation}\label{eqLagrangian}
\mathscr{L} = - \frac{1}{2}\bm{x}^{\intercal}W \bm{x} + \bm{\theta}^{\intercal}\bm{x} - \bm{\lambda}^{\intercal}\left(P \bm{x} - \bm{x}^{({\rm inc})}\right) + \frac{\gamma}{2} \bm{x}^{\intercal}\bm{x},
\end{equation}
where we introduce a Lagrange multiplier vector $\bm{\lambda} \in \mathbb{R}^{d}$ and a fixed regularization parameter $\gamma \geq 1$. The first-order derivative conditions for optimization are evaluated as
\begin{eqnarray}\label{Eq:FirstOrderLagrange}
\frac{\partial \mathscr{L}}{\partial \bm{x}} &=& (\gamma\mathbb{I}_{d} - W ) \bm{x} + \bm{\theta} - P \bm{\lambda} \stackrel{!}{=} 0 , \nonumber \\
\frac{\partial \mathscr{L}}{\partial \bm{\lambda}} &=& -P \bm{x} + \bm{x}^{({\rm inc})} \stackrel{!}{=} 0 .
\end{eqnarray}
One can equivalently consider this as a system of linear equations $A \bm{v} = \bm{w}$ with
\begin{eqnarray}\label{Eq:SystemOfEquations}
A &:=& \left( \begin{array}{cc}
W - \gamma \mathbb{I}_{d} & P \\
P & 0
\end{array}\right) ,\nonumber \\
\bm{v}&:=&
\left(
\begin{array}{c}
\bm{x} \\
\bm{\lambda}
\end{array}
\right)
, \qquad \bm{w}:=
\left(
\begin{array}{c}
\bm{\theta} \\
\bm{x}^{({\rm inc})}
\end{array}
\right).
\end{eqnarray}
The solution of this system then provides a vector $\bm{v}$ consisting of $\bm{x}$ and $\bm{\lambda}$, where $\bm{x}$ extremizes the energy $E$ subject to $P \bm{x} = \bm{x}^{({\rm inc})}$.
With $\Vert X \Vert$ the spectral norm (largest absolute eigenvalue) of a Hermitian matrix $X$, note from the definition in Eq.~(\ref{Eq:WeightingMatrix}) for the weight matrix $W$ that
$\Vert W \Vert \leq 1$. In addition, $\Vert \sigma_x \otimes P \Vert \leq 1$ and hence $\Vert A \Vert \in \mathcal O(\gamma)$. We set a reasonable choice of value for the regularization parameter to be $\gamma \in \mathcal O(1)$.  
It is shown in Appendix \ref{appConstrainedMin}
that the result of the optimization is necessarily a constrained local minimum of the energy whenever $\gamma$ is chosen such that 
$\gamma > \Vert W \Vert$.
Hence, it suffices to choose $\gamma > 1$. As the matrix $A$ is rank-deficient, we solve the system of equations by applying the pseudoinverse $A^{-1}$ to $\bm{w}$, recovering a least-squares solution to $\bm{v}$.

We find that the elements of the resultant vector $\bm{x}$ are continuous valued, i.e., $x_{i} \in \mathbb{R}$. 
This can be interpreted as a larger positive/negative value indicating a stronger confidence for the activation $\pm 1$, respectively. For a particular neuron, the value can then be projected to the nearest element $\pm 1$ to obtain a prediction for the activation of that neuron. The regularization term in the Lagrangian furthermore serves to minimize the $l_{2}$-norm $\left|\bm{x}\right|_{2}
$ of $\bm{x}$, and can be adapted by the user to prevent the optimization returning overly-large unconstrained elements, see Appendix \ref{appReg} for further details. Our approach to operating the Hopfield network through matrix inversion is tested in the Application section, using the example of RNA sequencing in genetics.

\section{ The quantum Hopfield network}

We now show how the Hopfield network can be run efficiently as a combination of quantum algorithms that we call qHop to perform the matrix inversion based approach. Utilizing the embedding method for quantum neural networks already discussed, the system of linear equations specified in~(\ref{Eq:SystemOfEquations}) can be written in terms of pure quantum states as $A \left|\bm{v}\right|_2\ket{v} = \left|\bm{w}\right|_2\ket{w}$, with $A$ as before, $P = \sum_{i \in \mathcal{L}}\ket{i}\bra{i}$, and
\begin{eqnarray}\label{Eq:QuantumSystemOfEquations}
\ket{v}&:=& \frac{1}{\left|\bm{v}\right|_2}\left(\left|\bm{x}\right|_2 \ket{0} \otimes \ket{x} + \left|\bm{\lambda}\right|_2 \ket{1}\otimes \ket{\lambda} \right),\nonumber \\
\ket{w}&:=& \frac{1}{\left|\bm{w}\right|_2}\left(\left|\bm{\theta}\right|_2 \ket{0} \otimes \ket{\theta} + |\bm{x}^{(\rm inc)}|_2 \ket{1}\otimes \ket{x^{(\rm inc)}} \right), \nonumber \\
\end{eqnarray}
being pure states of $N+1$ qubits.
Here, $\ket{x^{(\rm inc)}}$ is the normalized quantum state corresponding to the incomplete activation pattern and $|\bm{x}^{(\rm inc)}|_2^2 = l$. The objective is to optimize the energy function $E$ in Eq.~\eqref{Eq:HopfieldEnergy} by solving for $\bm{v} = \left|\bm{v}\right|_2 \ket{v} = A^{-1}\left|\bm{w}\right|_2\ket{w}$, with $A^{-1}$ the pseudoinverse of $A$.

It is possible to prepare $A^{-1}\ket{w}$ with a potential run time logarithmic in the dimension of $A$ by utilizing a combination of quantum subroutines. The objective is to use the quantum matrix inversion algorithm in Ref.~\cite{harrow2009quantum
}.  This algorithm requires the ability to perform quantum phase estimation using efficient Hamiltonian simulation of $A$. We now show that one can simulate $e^{i A t}$ by concurrently executing the simulation of a sparse Hamiltonian linked to the projector $P$ as well as qHeb. To achieve efficiency, certain conditions must be met. These conditions are outlined in the following sections.

We want to simulate the unitary $e^{i A t}$ to a fixed error $\epsilon$ for arbitrary $t$. Let us first write
\begin{eqnarray}
A &=& \left( \begin{array}{cc}
\rho - \left(\gamma+\frac{1}{d}\right) \mathbb{I}_{d} & P \\
P & 0
\end{array}\right) \nonumber \\
&=&
\left( \begin{array}{cc}
0 & P \\
P & 0
\end{array}\right)
+
\left( \begin{array}{cc}
-\gamma' \mathbb{I}_{d}  & 0 \\
0 & 0
\end{array}\right)
+
\left( \begin{array}{cc}
\rho & 0 \\
0 & 0
\end{array}\right) \nonumber \\
&=:& B + C + D,
\end{eqnarray}
where we introduce the $(2d \times 2d)$-dimensional block matrices
\begin{eqnarray}\label{Eq:LargerSpace}
B = \left( \begin{array}{cc}
0 & P \\
P & 0
\end{array}\right) \qquad \qquad C &=& \left( \begin{array}{cc}
-\gamma' \mathbb{I}_{d} & 0 \\
0 & 0
\end{array}\right) \nonumber \\
 \qquad D &=& \left( \begin{array}{cc}
\rho  & 0 \\
0 & 0
\end{array}\right)
\end{eqnarray}
with $\gamma' = \gamma + \frac{1}{d}$. We now split the simulation time $t$ into $n$
 small time steps $\Delta t$, i.e.~so that $t = n \Delta t$, and consider $e^{i A \Delta t}$. 
The time evolution $e^{i A \Delta t}$ can be simulated by using applications of $e^{i B \Delta t}$, $e^{i C \Delta t}$, and  $e^{i D \Delta t}$ via the standard Suzuki-Trotter method. Suppose that one has operators $\mathcal U_B(\Delta t) $, $\mathcal U_C(\Delta t) $, and $\mathcal U_D(\Delta t) $ that simulate $e^{i B \Delta t}$, $e^{i C \Delta t}$, and $e^{i D \Delta t}$ to errors at most $\mathcal{O}(\Delta t^2)$, respectively. In many cases much better error scalings exist. 
Then, $e^{i B \Delta t}e^{i C \Delta t}e^{i D \Delta t}$ is simulated to error also $\mathcal{O}(\Delta t^2)$. 
By simply using the Taylor expansion, we see that the error $\epsilon_{\Delta t}$ of simulating $e^{i A \Delta t}$ is
\begin{equation}
\epsilon_{\Delta t} := \left \Vert e^{i A \Delta t} - \mathcal U_B(\Delta t) \mathcal U_C(\Delta t)U_D(\Delta t) \right \Vert \in \mathcal{O}\left(\Delta t ^{2}\right).
\end{equation}
This means that by using $n$ repetitions of $\mathcal U_B(\Delta t) \mathcal U_C(\Delta t) \mathcal U_D(\Delta t)$ we can simulate $e^{i A t}$ to an error of $\epsilon \in \mathcal{O} \left(n \Delta t^{2}\right)$. 
Hence, for a fixed error $\epsilon$ and time $t$, one needs to perform $n \in \mathcal{O}\left(\frac{t^{2}}{\epsilon}\right)$ repetitions of $\mathcal U_B(\Delta t) \mathcal U_C(\Delta t)\mathcal U_D(\Delta t)$. 

We now evaluate the run time of performing one such repetition. Consider the block matrix $B$. Because $P$ is a diagonal projector, $B$ is a $1$-sparse self-adjoint matrix, where sparsity is the maximum number of elements in any column or row. A large series of works have addressed the efficient Hamiltonian simulation of sparse matrices. 
Reference~\cite{berry2015hamiltonian} shows that sparse Hamiltonian simulation for a simulation time $t$ to error $\epsilon$ can be performed with a run time $T_{B} \in \tilde {\mathcal{O}} (t \log (d)/\epsilon)$.
In our case, for the maximum matrix element of $B$ we have $\Vert B \Vert_{\max}=1$ and also $\Vert B\Vert =\mathcal{O}(1)$. The operator $\mathcal U_C(\Delta t)$ is treated in a similar way. Turning these operators $\mathcal U$ into their conditional versions and extending into a larger space as in Eq.~\eqref{Eq:LargerSpace} is in principle straightforward with the sparse matrix methods.  Simulating the operator $\mathcal U_D(\Delta t)$ is achieved using Hebbian learning, see Section \ref{secQHeb}, and including a conditioning on an additional ancilla qubit in state $\ket{0}$.

The essential steps of the algorithm are as follows and also summarized in Fig~\ref{Fig:Hopfield}. Let the spectral decomposition of $A$ be given by 
\begin{eqnarray}\label{Eq:ASplit}
A &=& \sum_{j : \,\, \left|\mu_{j}(A) \right| \geq \mu} \mu_{j}(A) \ket{v_j(A)} \bra{v_j(A)} \nonumber \\ & &\qquad + \sum_{j : \,\, \left|\mu_{j}(A)\right| <\mu}\mu_{j}(A) \ket{v_j(A)} \bra{v_j(A)},
\end{eqnarray}
where we have split into two separate sums dependent upon the size of the eigenvalues $\mu_{j}(A)$ in comparison to a fixed user-defined number $\mu > 0$. As we see in the following, as well as in Appendix \ref{appMu}, the chosen value of $\mu$ is a trade-off between the run time and the error in calculating the pseudoinverse. The primary matrix inversion algorithm returns (up to normalization)~\cite{harrow2009quantum}
\begin{equation}
A^{-1}\ket{w} = \sum_{j : \,\, \left|\mu_{j}(A)\right| \geq\mu} \frac{\beta_{j}}{\mu_j(A)}\ket{v_j(A)},
\end{equation}
where $\beta_{j} = \braket{v_{j}(A)|w}$. 

To begin, we first prepare the input state $\ket{w}$ (which contains the threshold data and incomplete activation pattern) and consider it in the eigenbasis of $A$, i.e. so that $\ket{w} = \sum_{j}\beta_{j} \ket{v_{j}(A)}$. Our qHeb algorithm is then initialized along with sparse Hamiltonian simulation~\cite{
berry2015hamiltonian} to perform quantum phase estimation, allowing us to obtain $\sum_{j} \beta_{j} \ket{\tilde{\mu}_{j}(A)}\otimes \ket{v_{j}(A)}$ with $\tilde{\mu}_{j}(A)$ an approximation of the eigenvalue $\mu_{j}(A)$ to precision $\epsilon$. We then use a conditional rotation of an ancilla and a filtering process discussed in Ref.~\cite{harrow2009quantum} to select only the eigenvalues larger than or equal to $\mu$. This is followed by an uncomputing of the first register of $T$ qubits by reversing the quantum phase estimation protocol. After measurement of the ancilla qubit, our result is (up to normalization) the pure state $A^{-1}\ket{w}$. 

A note regarding the input state $\ket{x^{(\rm inc)}}$. In principle, for each reconstruction of a new input state, we require new runs of qHeb and qHop. This feature arises from the no-cloning theorem for quantum states. Different from classical computing, one in general cannot efficiently copy intermediate data of single runs of the algorithm for reuse to reconstruct other input patterns. However, one can envision scenarios where one can reconstruct multiple patterns simultaneously via a quantum superposition of the input patterns. Let $\ket{x^{(\rm inc,k)}}$, $k=1,\dots,K$ be $K$ patterns. Assume we can prepare superpositions of the form
$\ket{x^{(\rm inc,total)}} = \sum_{k=1}^K \alpha_k \ket{x^{(\rm inc,k)}}$ or $\ket{x^{(\rm inc,total)}} = \sum_{k=1}^K \alpha_k \ket k \ket{x^{(\rm inc,k)}}$, with coefficients $\alpha_k$ such that the total state is normalized in each case and $\ket k$ a label register. Then we can use the qHop algorithm by replacing $\ket{x^{(\rm inc)}}$ by $\ket{x^{(\rm inc,total)}}$. We then are able to extract information about the $K$ patterns from the resulting state, see the discussion of the output state in Sec.~\ref{sectionOutput}. Of course obtaining information on each individual pattern will again require $ \mathcal{O}\left(K \right)$ operations of qHop, but we can hope to extract summary statistics with fewer resources.

\section{Algorithm Efficiency}

We now turn to addressing the efficiency of qHop. The overall efficiency is not just dependent upon the run time of our primary algorithm, and we must also consider the read-in efficiency of inputting $\ket{w}$ as well as the read-out efficiency of extracting useful information from the output state $\ket{v}$. Here we review the input and run-time efficiencies, while the next section discusses various ways of using the output and their efficiency. The section after briefly  compares our qHop to other classical and quantum approaches to operating the Hopfield network.

The input pure state $\ket{w}$ contains data on the user-specified neuronal thresholds $\bm{\theta}$, along with the incomplete activation pattern $\bm{x}^{(\rm inc)}$. As we have discussed, the read-in of activation patterns can add a computational overhead to quantum neural network algorithms, potentially canceling any speed-ups yielded by the algorithm itself. This can be addressed using, e.g., qRAM~\cite{giovannetti2008quantum} or efficient state preparation techniques~\cite{soklakov2006efficient}
, or alternatively by directly accessing the output of a quantum device. Let us denote by $T_{\rm in}$ the run time of inputting $\ket{w}$, which we take to be $\mathcal{O}\left({\rm poly} \left(\log d \right)\right)$ using any of the discussed techniques. Note that state preparation techniques may introduce errors themselves, but these can be fixed to $\epsilon$ and will typically add a polynomial overhead in $\epsilon$ to the run-time~\cite{soklakov2006efficient}.

Following similar calculations to those discussed in Ref.~\cite{harrow2009quantum}, we see that our algorithm proceeds by a combination of phase estimation of $A$ with run time $T_{\rm phase}$ along with filtering and amplification operations to select the eigenvalues $\vert \lambda_{j}(A) \vert \geq \mu$~\cite{harrow2009quantum}, requiring a run time $T_{\rm filter}$. Let us consider first phase estimation, which requires us to perform $\mathcal{O}\left(\frac{1}{\epsilon^{3}}\right)$ calls to $e^{i A t}$. One can decompose $A$ into three block matrices $B$, $C$, and $D$, corresponding to the off-diagonal projector $P$, an on-diagonal identity $\mathbb{I}_{d}$, and, when using Hebbian learning, the embedded mixed training state $\rho$, see Eq.~\eqref{Eq:SystemOfEquations}. As we have shown, $e^{i A t}$ is well approximated by applying for $n$ short times $\Delta t$ the unitaries $U_{B/C/D}$ generated by these block matrices, resulting in an error $\epsilon \in \mathcal{O}\left(\frac{t ^{2}}{n} \right)$
or equivalently requiring a number of steps $n = \mathcal{O}\left(t^{2}/\epsilon\right)$. 

Since both $B$ and $C$ are $1$-sparse matrices, we can use efficient sparse Hamiltonian simulation techniques~\cite{
berry2015hamiltonian} to evaluate $U_{B/C}(\Delta t)$ with run time $T_{B/C} \in \mathcal{ O}\left( {\rm poly} \left( \Delta t,\log d , \log \left(\frac{1}{\epsilon}\right)\right)\right)$. For the matrix $D$, we can use the quantum Hebbian learning techniques discussed earlier 
to simulate for a time $\Delta t$, requiring a run time $T_{D} \in \mathcal{O}\left(  {\rm poly}\left(\Delta t, M, \frac{ 1}{\epsilon}, \log d\right)\right)$. 
Note that the state exponentiation technique used for $D$ means that $T_{D}$ is the dominant run time compared to $T_{B/C}$. 
Hence, overall we have $T_{\rm phase} \in \mathcal{O}\left({\rm poly}\left(M,\log d , \frac{1}{\epsilon} \right)\right)$. The run time for filtering and amplification adds an additional overhead $T_{\rm filter } \in \mathcal{O}\left( \frac{1}{\mu} \right)$~\cite{harrow2009quantum}
, meaning that the user should set $1/\mu \in \mathcal{O} \left({\rm poly}\left(\log{d} \right)\right)$ to maintain efficiency. 
We hence achieve an overall algorithm run time of
\begin{equation}\label{Eq:AlgEff}
T_{\rm qHop} \in \mathcal{O}\left({\rm poly}\left(M,\log d , \frac{1}{\epsilon} , \frac{1}{\mu} \right)\right).
\end{equation}

A note on the $M$ dependence. The maximum capacity of the classical Hopfield network is approximately $d/(2 \log d)$~\cite{mceliece1987capacity} memory patterns. The linear dependence on $M$ of the quantum algorithm means that for achieving a logarithmic dependency on the dimension, qHop has to be operated substantially below the maximum capacity. Any potential exponential speedup arises from the processing of these $d$-dimensional memory patterns, while the number of the memory patterns has to be relatively small.
To extend the range when one may observe speedups, we can consider a scenario when the density weight matrix is directly given and we can use the original density matrix exponentiation scheme \cite{lloyd2013quantum,kimmel2017hamiltonian}. This scenario does not require our Hebbian learning and avoids the $M$ dependence. Moreover, in the case when the weight matrix is given via oracle access to the matrix elements and is sparse, one use the sparse simulation techniques \cite{berry2015hamiltonian}. In this case, we can directly use qHop without requiring the Hebbian learning procedure and the $M$ dependence is absorbed into the oracle.

The output of our algorithm is the pure state $\ket{v}$ given in Eq.~(\ref{Eq:QuantumSystemOfEquations}). We can then measure the first qubit in our $N+1$ qubit register and post-select on $\ket{0}$ to obtain $\ket{x}$. This succeeds with probability $\left| \bm{x} \right|_2^{2}/(\left| \bm{x} \right|_2^{2}+\left| \bm{\lambda} \right|_2^{2})$, adding a processing overhead $T_{\rm ps} \in \mathcal{O}\left({\left| \bm{\lambda} \right|_2^{2}}/{\left| \bm{x} \right|_2^{2}}\right)$. One can see from Eq.~\eqref{Eq:FirstOrderLagrange} that $x_{i} \in \mathcal{O}\left(1\right)$ for the constrained neurons $i \in \mathcal{L}$ and $x_{i} \in \mathcal{O}\left(\frac{1}{\gamma}\right)$ for the unconstrained neurons, so that $\left| \bm{x} \right|_2^{2} \in \mathcal{O}\left(d\right)$ whenever the number of constrained neurons $l$ is of the order $d$. On the other hand, since $\lambda_{i}\in \mathcal{O}\left(\gamma\right)$ for $i \in \mathcal{L}$ and $\lambda_{i} =0$ otherwise, we have $\left| \bm{\lambda} \right|_2^{2} \in \mathcal{O}\left(d \gamma^{2}\right)$. Hence, overall our processing overhead is $T_{\rm ps} \in \mathcal{O}\left(\gamma^{2}\right)$. This means that our choice of $\gamma$ is in fact a compromise, one must pick $\gamma \geq \Vert W \Vert$ to guarantee a local minimum, but if $\gamma$ is too large then we add a run-time overhead to qHop.
The next section discusses what to do with the output state at what cost to the efficiency.

\section{Output}
\label{sectionOutput}

The next step is naturally to use the information contained in $\ket x$ for a given task. One way to use the state is to read-out the amplitudes of $\ket{x}$ by performing tomography. However, even for pure states, tomographical techniques can introduce an overhead that scales polynomially with the dimension $d$~\cite{
gross2010quantum}. Instead, one has to extract useful information from $\ket{x}$ using other approaches, which typically act globally on $\ket{x}$ rather than directly accessing each of the $d$ amplitudes. 
Such extraction of global information aligns well with typical situations in machine learning. Machine learning tasks often involve dimensionality reduction or compression. For example an image of many pixels is compressed to a single label (`cat' or `dog') or a short description of the scene in that image. Classifications tasks often involve a small number of classes, for example users of a movie streaming service can be assigned to a relatively small number of categories \cite{Kerenidis2016}. In the context of neural networks, both artificial and biological, the state of a single intermediate neuron is rarely important to a  learning task, but rather the final goal is to obtain a low-dimensional explanation or action which relies on the output patterns of a larger collection of neurons. 

One option to extract global information could be to measure the fidelity with another state $\ket{\tilde{x}}$, such as one of the training states, which can be achieved by performing a swap test with success probability $P_{\rm swap}= \frac{1}{2} \left( 1- \left|\braket{\tilde{x}|x}\right|^{2} \right)$~\cite{gottesman2001quantum}. We can then determine the fidelity to a precision $\epsilon$ by performing $\mathcal{O}\left(\frac{P_{\rm swap}(1-P_{\rm swap})}{\epsilon^2}\right)$ swap tests between copies of $\ket{x}$ and $\ket{\tilde{x}}$, with each swap test requiring $\mathcal{O}\left(\log d\right)$ qubit swaps and hence giving an additional run time to qHop of $T_{\rm out} \in \mathcal{O}\left({\rm poly} \left(\log d , \frac{1}{\epsilon}\right)\right)$.

Alternatively, following the spirit of supervised learning, one may have access to a set of $p$ binary valued observables, corresponding to membership of some classification categories. Measuring the expectation values of these observables with respect to $\ket{x}$ then allows for a classification of $\ket{x}$ with respect to such categories. For a given precision $\epsilon$, each expectation value can be measured with $\mathcal{O}\left(\frac{1}{\epsilon^{2}}\right)$ repetitions, resulting in a run-time overhead to qHop of $T_{\rm out} \in \mathcal{O}\left({\rm poly} \left(\frac{1}{\epsilon},p,T_{\rm obs}\right)\right)$, with $T_{\rm obs}$ the time of the observable measurement.

In addition, one can adopt a fully quantum perspective and view the state $A^{-1}\ket{w}$ (or the post-selected activation pattern state $\ket{x}$), as the final output of the algorithm. Our qHop algorithm then acts as an element of a given quantum toolchain, whose action is to reconstruct a quantum state from an incomplete superposition based on the memory stored in $\rho$, and then to output to the next element in the chain.

\section{Comparison}
To summarize, the full operation of qHop can be achieved with a run time $\mathcal{O}\left({\rm poly}\left(M, \log d , \frac{1}{\epsilon}  , \frac{1}{\mu} \right)\right)$, where Fig.~\ref{Fig:Hopfield} visualizes the individual run time contributions. 
We now compare this efficiency with both of the classical approaches: the original Hopfield procedure~\cite{hopfield1982neural}, as well as the new matrix inversion based approach introduced here. It is clear that the original Hopfield procedure has a run time polynomial in the number of neurons, since one must typically sample every one of the $d$ neurons at least once. On the other hand, the best  sparse classical matrix inversion techniques have a run time $\mathcal{O}\left({\rm poly} \left(d , \frac{1}{\sqrt{\mu}} , \log \left(\frac{1}{\epsilon}  \right), s \right)\right)$~\cite{shewchuk1994introduction} where $s$ is the sparsity, and it has been shown in Ref.~\cite{harrow2009quantum} that this run time cannot be improved even if one needs access only to the expectation values of $A$. We hence see that qHop is potentially able to operate with lower computational demands for a suitably large $d$. 
Of course, better classical algorithms can be found for example harnessing the similarities between the Hopfield network and the Ising model that is studied in-depth in quantum physics~\cite{dunjko2017machine,ising1925beitrag}. Techniques such as simulated annealing~\cite{van1987simulated} and mean field theory~\cite{opper2001advanced} can help provide a better account of classical performances.

We briefly compare with other quantum approaches. 
The first quantum Hopfield network \cite{ventura1998quantum} encodes the data in the basis states of an exponentially large quantum state (instead of using the amplitudes) and uses Grover search to attain quantum speedups for memory recall. Such Grover search speedups are possible in rather generic settings, achieving a performance of about
$\sqrt{d}$. 
In the adiabatic quantum computing framework, quantum Hopfield networks are developed, exploiting the natural connection of the Hopfield network  and Ising-like energy functions \cite{seddiqi2014adiabatic}. The critical quantity for the run time is the spectral gap of the associated Hamiltonian. In many cases, this spectral gap is exponentially small, leading to similar run-times as the classical methods. In other cases, when the gap is only polynomially small, exponential speedups may be possible.
Reference~\cite{rotondo2017open} considers an open quantum system treatment of the Hopfield network and develops the resulting phase diagram. Quantum effects are shown to be included by an effective temperature.
Other works \cite{akazawa2000quantum} have discussed single-electron quantum tunneling in the context of the Hopfield network, which can overcome local energy minima, where actual performance will be determined by the physical implementation. Another work has discussed the potential occurrence of quantum effects in cellular microtubules at low temperatures \cite{behrman2006microtubules}.

Our work uses an exponential encoding of the neuronal information into quantum amplitudes, the gate model of quantum computing, and a setting where quantum phase estimation and matrix inversion can be used for the Hopfield network. As discussed these techniques can lead to potential performance logarithmic in the number of neurons for specific applications.
However, let us emphasize that this analysis does not constitute a comprehensive benchmark of qHop against possible classical and quantum approaches to running the Hopfield network.

\section{Application}

\begin{figure*}
  \begin{center}
  \includegraphics[width=\textwidth]{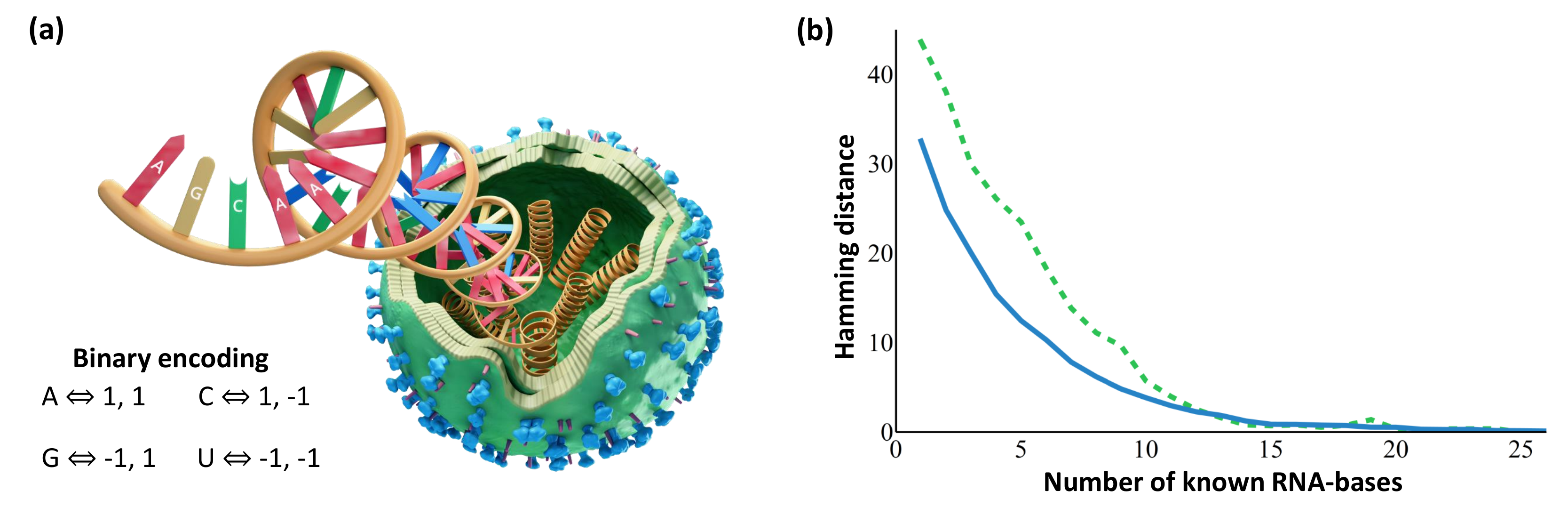}
\end{center}
    \vspace{-1em}
\caption{
RNA recognition. {\bf (a)} The Hopfield network can be used as a content addressable memory system for RNA-recognition (data source \cite{zaraket2010genetic}). We encode $50$ RNA-bases of the $M=8$ strands originating from the H1N1 influenza A virus in $W$, and then run the Hopfield network on partial information from a limited number of randomly selected RNA-bases from the first strand. {\bf (b)} The result of operating the Hopfield network on this example using the standard classical approach (dotted line) and the matrix inversion based approach (solid line). The resultant Hamming distance to the true data is averaged over $1000$ repetitions for varying amounts of partial information.
}
  \label{Fig:Flu}
\end{figure*}

Here we outline an application of the Hopfield network in RNA sequencing. Consider the H1N1 strain of the influenza A virus, which has 8 RNA segments that code for different functions in the virus. The segments are composed of a string of RNA-bases: A, C, G, and U. Each segment can in turn be converted to a double sized binary string, as shown in Fig.~\ref{Fig:Flu},
which can be stored in the weighting matrix of a Hopfield network. Suppose that we are provided with partial information on a new RNA sequence and would like to verify whether it belongs to the H1N1 virus. For example, our sequence could be from a recently collected sample originating in an area with a new influenza outbreak. This scenario can be addressed by resorting to the Hopfield network.

We use this setting as a motivation for our numerics presented in Fig.~\ref{Fig:Flu}, which contains a comparison of the performance of the standard classical approach to operating the Hopfield network with our new matrix inversion based approach. Here, we store the first $50$ RNA-bases from each of the $8$ segments of the influenza A H1N1 strain (i.e. so that $d=100$, $M=8$) in the weighting matrix $W$ using the Hebbian learning rule (data source~\cite{zaraket2010genetic}). For this small example, the weighting matrix is filled to 
classical capacity, i.e., $M=5 \approx d/(2 \log d)$~\cite{mceliece1987capacity}, so that imperfect recoveries are more easily identified. Note the discussion on the $M$ dependency after Eq.~(\ref{Eq:AlgEff}). We then generate incomplete data from the first segment of H1N1 by randomly selecting $l/2$ RNA-bases for $l/2 \in \{1,2,\ldots,50\}$. Both approaches to operating the Hopfield network are then implemented to reconstruct the full activation pattern, with the Hamming distance measured between the result and the original pattern. This is averaged over $1000$ repetitions of random choices of $l/2$ RNA-bases, with the resultant data plotted in Fig.~\ref{Fig:Flu}. We see that both the conventional approach to the Hopfield network and the new matrix inversion based approach have comparable performances, with each able to recover the input segment for a suitably large $l/2$. Yet, by using qHop to perform the matrix inversion based approach, we could operate with a run time logarithmic in the system dimension and hence increase the dimension far beyond $d=100$, see the previous section and Fig.~\ref{Fig:Hopfield} for a comparison of run times. Note that for the matrix inversion based approach, we set $\gamma = 1$ to guarantee a local minimum since $\Vert W\Vert \approx 0.185$. Moreover, the objective is to classify whether the collected sample is the H1N1 virus. For the quantum version of the Hopfield network, this can be achieved by performing a swap test with the target state $\ket{\tilde{x}}$ set to correspond to the encoded H1N1 virus.

\section{DISCUSSION}

Quantum effects have a profound potential to yield advancements in machine learning over the coming decade. We have presented a quantum implementation (qHop) for the Hopfield network that encodes an exponential number of neurons within the amplitudes of only a polynomially large register of qubits. This complements alternative encodings focusing on a one-to-one correspondence between neurons and qubits. Crucially, the learning and operation steps of the quantum Hopfield network can be exponentially quicker in run time when compared to classical approaches. We have also introduced a method of training a quantum neural network via quantum Hebbian learning (qHeb).

As with many quantum algorithms, the efficient operation of qHop is subject to some important considerations. One must first be able to efficiently read-in the classical initialization data of the neural network into our quantum device, which can be achieved using efficient pure state preparation techniques~\cite{
soklakov2006efficient} or qRAM~\cite{giovannetti2008quantum}, or alternatively by directly using the output of a quantum device. Next, it must be possible to operate efficiently qHeb, and matrix inversion~\cite{harrow2009quantum}. This relies on efficient Hamiltonian simulation of the system matrix, which we show to be possible by resorting to sparse Hamiltonian simulation techniques~\cite{
berry2015hamiltonian} and density matrix exponentiation~\cite{lloyd2013quantum,kimmel2017hamiltonian}. 
When using qHeb as a learning method, we obtain a linear dependence on the number of training examples, which affects the capacity of the quantum Hopfield network. This linear dependence  may be avoided by using sparse simulations or density matrix simulation directly on the qHop matrix.
The matrix inversion algorithm then outputs the inverse only on a well-conditioned subspace with (absolute) eigenvalues larger than a chosen fixed value $\mu$ whose inverse controls the algorithm efficiency. It is crucial to note that classical sparse matrix inversion algorithms also have a similar efficiency-dependence on $\mu$. Finally, it must be possible to efficiently access the output of qHop, which is a pure quantum state representing a continuous-valued neuronal activation pattern. Since a quantum state tomography is typically resource intensive, one can instead access global information such as the fidelity with previously trained activation patterns or the expectation values with respect to observables.

We have introduced the subroutine qHeb, which adapts the standard Hebbian learning approach~\cite{hebb1949organization} to the quantum setting, a new addition to studies on quantum learning. 
 Our subroutine relies on the important observation that the weight matrix $W$ describing a neural network can be alternatively represented by a mixed quantum state (or more generally, a Hamiltonian). Using density matrix exponentiation~\cite{lloyd2013quantum,kimmel2017hamiltonian}, this quantum state can then be used operationally for the extraction of, e.g., eigenvalues and eigenvectors of the weight matrix. We have shown that quantum Hebbian learning can be implemented by performing a sequential imprinting of memory patterns, represented as pure quantum states, onto a register of memory qubits. 
Although introduced here within the context of the quantum Hopfield network, quantum Hebbian learning 
can be of wider interest as a quantum subroutine within other quantum neural networks.

Our findings, along with other work~\cite{kak1995quantum,bonnell1997quantum,altaisky2001quantum,narayanan2000quantum,schuld2014quest,ezhov2000quantum,wan2016quantum,amin2016quantum,wiebe2014quantum,kieferova2016tomography,benedetti2017quantum}, including quantum Hopfield networks~\cite{behrman2000simulations,akazawa2000quantum,behrman2006microtubules,rotondo2017open}, contribute to the goal of developing a practical quantum neural network. The approach we use encodes an exponential number of neurons into a polynomial number of qubits. We have discussed a specific neural network, the Hopfield network, which is a content addressable memory system. 
As an application, we have shown how the matrix inversion-based Hopfield network can be utilized for identifying genetic segments of RNA in viruses. Future developments may focus on the nature of quantum neural networks themselves, identifying entirely new applications that harness purely quantum properties without being based upon previous classical networks. The natural next step to benefit from the fruits of quantum neural networks, and developments in quantum machine learning more generally, is to implement these algorithms on near-term quantum devices.

\begin{acknowledgements}
We thank Juan Miguel Arrazola, Mayank Bhatia and Nathan Killoran for fruitful discussions.  S.~L. was supported by OSD/ARO under the Blue Sky Initiative. 
\end{acknowledgements}

\bibliographystyle{apsrev}
\bibliography{References}

\appendix

\section{Perturbed data}
\label{appendixNoisy}
Instead of incomplete data, we here discuss the problem of correcting perturbed data. The new element is 
$\bm{x}^{({\rm new})}$ and its perturbed version is $\bm{x}^{({\rm pert})}$. We pose the classical problem by including the closeness to the perturbed version via an l2-norm constraint. Instead of the Lagrangian Eq.~(\ref{eqLagrangian}), we have the error function
\begin{equation}
\mathscr{E} := - \frac{1}{2}\bm{x}^{\intercal}W \bm{x} + \bm{\theta}^{\intercal}\bm{x} + \frac{\beta}{2} \left\vert \bm{x} - \bm{x}^{({\rm pert})}\right\vert_2^2 + \frac{\gamma}{2} \bm{x}^{\intercal}\bm{x},
\end{equation}
where instead of  Lagrange multipliers we use the regularization parameter $\beta$. 
The first-order criterion now is
\begin{eqnarray}
\frac{\partial \mathscr{E}}{\partial \bm{x}} &=& ((\gamma + \beta)\mathbb{I}_{d} - W ) \bm{x} + \bm{\theta} - \beta \bm{x}^{({\rm pert})} \stackrel{!}{=} 0. 
\end{eqnarray}
This leads to the matrix inversion problem for finding $\bm{x} $ as
\begin{equation}
\left ( (\gamma + \beta)\mathbb{I}_{d} - W \right) \bm{x} =   \beta \bm{x}^{({\rm pert})} -\bm{\theta}.
\end{equation}
The resulting quantum algorithm is similar, even slightly simpler, than the method discussed in the main part of this work, and not further discussed here.

\section{Constrained Minimization of the Energy Function}
\label{appConstrainedMin}

Here we show that the result of the constrained optimization outlined in the Hopfield network section of the main text is necessarily a local minimum. Suppose that we are to optimize a real-valued scalar function $f(\bm{x})$ of a vector $\bf{x}\in \mathbb{R}^{d}$ subject to $l < d$ constraints composed into a real-valued vector function $\bf{g}(\bf{x}) = \bf{0}$. The corresponding Lagrangian is $\mathscr{L}(\bm{x},\bm{\lambda}) = f(\bm{x}) - \bm{\lambda}^{\intercal}\bf{g}(\bf{x})$ with Lagrange multiplier vector $\bm{\lambda}$. Optimization can be achieved by identifying vectors $(\tilde{\bm{x}},\tilde{\bm{\lambda}})$ satisfying $\partial_{\bm{x}} \mathscr{L} = \bm{0}$ and $\partial_{\bm{\lambda}} \mathscr{L} = \bm{0}$. To classify these optimal vectors we must consider the $((l+d)\times (l+d))$-dimensional bordered Hessian matrix~\cite{ghosh2011polynomial}
\begin{equation}
\mathscr{H}(\bm{x},\bm{\lambda}) := \left( \begin{array}{cc}
0_{l} & \nabla \bf{g}(\bf{x}) \\[6pt]
\nabla \bf{g}(\bf{x})^{\intercal} & \frac{\partial^{2} \mathscr{L}}{\partial \bm{x}^{2}}
\end{array}\right).
\end{equation}
In particular, $(\tilde{\bm{x}},\tilde{\bm{\lambda}})$ is a local minimum if
\begin{equation}
(-1)^{l} {\rm det} \left(\mathscr{H}_{k}(\bm{x},\bm{\lambda})\right) > 0
\end{equation}
for all $k \in \{2l+1,2l+2,\ldots,l+d\}$, where $\mathscr{H}_{k}(\bm{x},\bm{\lambda})$ is the $k$-th order leading principle submatrix of $\mathscr{H}(\bm{x},\bm{\lambda})$, composed of taking the first $k$ rows and the first $k$ columns.

We now show that this condition is satisfied when $f(\bm{x})$ is the energy $E$ in Eq.~(\ref{Eq:HopfieldEnergy}) of the main text and $\bf{g}(\bf{x}) = P \bm{x} - \bm{x}^{({\rm inc})}$. The bordered Hessian matrix is then
\begin{equation}
\mathscr{H}(\bm{x},\bm{\lambda}) = \left( \begin{array}{cc}
0_{l} & -\tilde{P} \\[6pt]
-\tilde{P}^{\intercal} & \gamma \mathbb{I}_{d} - W
\end{array}\right),
\end{equation}
with $\tilde{P}$ a rectangular $(l \times d)$-dimensional matrix of rows of unit vectors $e_{i}$ for $i \in \mathcal{L}$, or equivalently the projector $P$ with all zero rows removed. We note that in our setting the bordered Hessian matrix is in fact independent of $\bm{x}$ and $\bm{\lambda}$, meaning that we can classify any extremum found. We therefore herein drop the following brackets around $\mathscr{H}$. Now consider the leading principle minor $\mathscr{H}_{k}$ for any $k \in \{2l+1,2l+2,\ldots,l+d\}$, given by
\begin{equation}
\mathscr{H}_{k} = \left( \begin{array}{cc}
0_{l} & -\tilde{P}_{l \times (k-l)} \\[6pt]
-(\tilde{P}_{l \times (k-l)})^{\intercal} & (\gamma \mathbb{I}_{d} - W)_{(k-l)}
\end{array}\right),
\end{equation}
with $\tilde{P}_{l \times (k-l)}$ composed of the first $k-l$ columns of $\tilde{P}$ and $(\gamma \mathbb{I}_{d} - W)_{(k-l)}$ the $(k-l)$-th order leading principal submatrix of $\gamma \mathbb{I}_{d} - W$. 

\begin{figure}[t!]
  \begin{center}
  \includegraphics[width=0.48\textwidth]{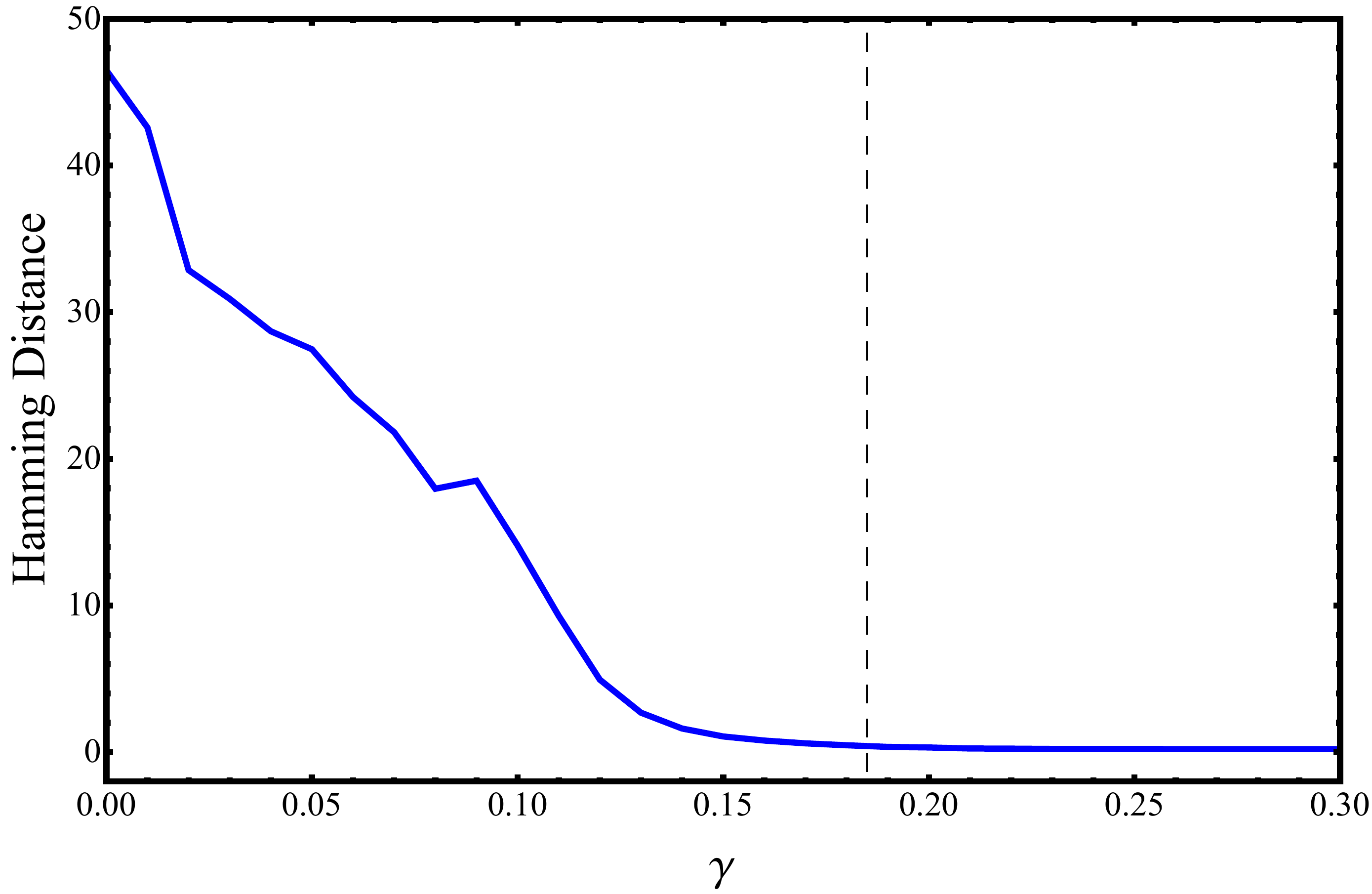}
\end{center}
    \vspace{-1em}
\caption{The average Hamming distance between one of the reconstructed memory patterns and the original when $l=50$ neurons are known \emph{a priori}, given as a function of the regularization parameter $\gamma$. The maximum eigenvalue $\left|\left|W\right|\right|\approx 0.185$ of $W$, which $\gamma$ must exceed to guarantee a local minimum, is shown as the vertical dashed line. Note that no increase of the Hamming distance is observed for $\gamma > 0.3$.
}
  \label{Fig:Reg}
\end{figure}

Let us consider $\gamma > \Vert W\Vert$ with $\Vert W\Vert$ the largest eigenvalue of $W$, so that $\gamma \mathbb{I}_{d} - W > 0$. Sylvester's criterion tells us that $(\gamma \mathbb{I}_{d} - W)_{(k-l)} > 0$ and is hence invertible. Using the Schur complement, we have that
\begin{eqnarray}
{\rm det}\left(\mathscr{H}_{k}\right) &=& (-1)^{l}{\rm det}\left((\gamma \mathbb{I}_{d} - W)_{(k-l)}\right)\\
&&\times \,\, {\rm det}\left( \tilde{P}_{l \times (k-l)}\left((\gamma \mathbb{I}_{d} - W)_{(k-l)}\right)^{-1} \right. \nonumber \\ && \qquad \qquad \times \,\, \left. (\tilde{P}_{l \times (k-l)})^{\intercal} \right), \nonumber
\end{eqnarray}
with $X^{-1}$ the inverse of $X$. On the other hand, we know that $\left((\gamma \mathbb{I}_{d} - W)_{(k-l)}\right)^{-1} >0$. The action of $\tilde{P}_{l \times (k-l)}\left((\gamma \mathbb{I}_{d} - W)_{(k-l)}\right)^{-1}(\tilde{P}_{l \times (k-l)})^{\intercal}$ is to select an $l$-th order principal minor of $\left((\gamma \mathbb{I}_{d} - W)_{(k-l)}\right)^{-1}$. It is a well known result in linear algebra that any principle minor of a positive definite matrix is itself positive definite~\cite{shilov1977linear}, so that we know $\tilde{P}_{l \times (k-l)}\left((\gamma \mathbb{I}_{d} - W)_{(k-l)}\right)^{-1}(\tilde{P}_{l \times (k-l)})^{\intercal}>0$ for any $k$. Since the determinant of a positive definite matrix is positive, we hence know that
\begin{eqnarray}
{\rm det}\left((\gamma \mathbb{I}_{d} - W)_{(k-l)}\right) &>& 0, \nonumber \\
{\rm det}\left( \tilde{P}_{l \times (k-l)}\left((\gamma \mathbb{I}_{d} - W)_{(k-l)}\right)^{-1}(\tilde{P}_{l \times (k-l)})^{\intercal} \right) &>& 0. \nonumber
\end{eqnarray}
This means that the sign of ${\rm det} \left(\mathscr{H}_{k}\right)$ is given by $(-1)^{l}$, and that overall
\begin{equation}
(-1)^{l} {\rm det} \left(\mathscr{H}_{k}\right) > 0,
\end{equation}
satisfying the condition for a minimum given above. \\

\section{Setting the Regularization Parameter}
\label{appReg}

From the previous section, we see that it is necessary to introduce the regularization parameter to provide a sufficient condition that our constrained optimization reaches a local minimum. From the perspective of machine learning, the regularization parameter also functions to penalize large values of $\left|\bm{x}\right|_{2}$ in the minimization to prevent over-fitting. In Fig.~\ref{Fig:Reg}, following the example outlined in the main text, we plot the average Hamming distance between the reconstructed pattern (using our matrix-inversion based approach with discretized post processing) and the original pattern for increasing values of regularization parameter and a constant number of known neurons $l=50$. Here, the average Hamming distance drops off dramatically to zero for a sufficiently high regularization parameter $\gamma > \Vert W \Vert \approx 0.185$. However, if one chooses an arbitrary large $\gamma$ then this adds a polynomial run time onto qHop (see the efficiency discussion in the main text). In the numerics of the main part, we set $\gamma = 1$.

\section{Setting the value of \texorpdfstring{$\mu$}{mu}}
\label{appMu}

Our algorithm finds the inverse of
\begin{equation}
\tilde{A}:=\sum_{j : \,\, \left|\mu_{j}(A) \right| \geq \mu} \mu_{j}(A) \ket{v_j(A)} \bra{v_j(A)},
\end{equation}
[see Eq.~(\ref{Eq:ASplit}) of the main text for comparison to $A$.] It holds that $\tilde{A}^{-1}$ is equal to the pseudoinverse $A^{-1}$ whenever $\mu$ does not exceed the smallest nonzero singular value $\left|\mu_{\min}\right|$ of $A$. Otherwise, $\tilde{A}^{-1}\ket{w}$ approximates $A^{-1}\ket{w}$ to an error
\begin{equation}
\eta := \left \vert \tilde{A}^{-1}\ket{w} - A^{-1}\ket{w}\right \vert_2.
\end{equation}
From Eq.~(\ref{Eq:AlgEff}) of the main text, it can be seen that qHop maintains the polylogarithmic efficiency in run time whenever $\mu$ is such that $1/\mu \in {O}\left({\rm poly}\left(\log{d}\right)\right)$. Hence, for the matrix inversion to be effective, we require $A$ to be such that either (1) $\left|\mu_{\min}\right| \geq \mu$, with no additional errors in finding the pseudoinverse, or (2) $\left|\mu_{\min}\right| < \mu$ but with $\eta \in {O}\left(\epsilon \right)$ so that the error $\eta$ accumulates in accordance with an overall desired error $O(\epsilon)$.

\end{document}